\documentclass[%
nofootinbib, 
 aps,
 prl,%
reprint,%
]
{revtex4-1}

\usepackage{}
\usepackage{xr}
\usepackage[normalem]{ulem}

\usepackage{bm}
\usepackage[pdftex]{graphicx}
\usepackage{amsmath,amsfonts,amsthm,amssymb,latexsym,mathrsfs}
\usepackage{multirow} 
\usepackage{tabularx}
\renewcommand\figurename{Figure}

\begin{document}
  \makeatletter
  \renewcommand\@biblabel[1]{#1. }
  \makeatother
  \makeatletter
  \renewcommand{\fnum@figure}[1]{\textbf{\figurename~\thefigure}.}
  \makeatother 
  
\title{Shape Transitions and Chiral Symmetry Breaking in the Energy Landscape of the Mitotic Chromosome}

\author{Bin Zhang$^{a,b}$}
\affiliation{
Department of $^a$Chemistry and $^c$Physics and Astronomy, 
and $^{b}$Center for Theoretical Biological Physics, Rice University, Houston, TX 77005} 
\author{Peter G. Wolynes$^{a,b,c}$} 
\email[Correspondence: ]{pwolynes@rice.edu}
\affiliation{
Department of $^a$Chemistry and $^c$Physics and Astronomy, 
and $^{b}$Center for Theoretical Biological Physics, Rice University, Houston, TX 77005}


\begin{abstract}
We derive an unbiased information theoretic energy landscape for chromosomes at metaphase using a maximum entropy approach that accurately reproduces the details of the experimentally measured pair-wise contact probabilities between genomic loci. Dynamical simulations using this landscape lead to cylindrical, helically twisted structures reflecting liquid crystalline order. These structures are similar to those arising from a generic ideal homogenized chromosome energy landscape. The helical twist can be either right or left handed so chiral symmetry is broken spontaneously. The ideal chromosome landscape when augmented by interactions like those leading to topologically associating domain (TAD) formation in the interphase chromosome reproduces these behaviors. The phase diagram of this landscape shows the helical fiber order and the cylindrical shape persist at temperatures above the onset of chiral symmetry breaking which is limited by the TAD interaction strength.

\end{abstract}

\maketitle

When cells divide, their chromosomes dramatically condense before forming a pair of sister chromatids that exhibits the famous X shape of the mitotic phase \cite{Lod08}. Microscopy reveals the overall morphology of the mitotic chromatin, but its internal structure remains controversial \cite{Mar08, Kir04, Mae03}. A complete understanding of chromosomal organization is challenging since the condensation of the chromosome into its dense mitotic form permits many different orderings and phase transitions. Owing to the large scale of the chromosome, we also can expect interesting non-equilibrium glassy dynamics and possible kinetic control of structure formation \cite{Zha15, Kan15, Lie09, Nau13, Ste13, Boh10, Bar12, Gur14, Ben13}. 

Structural models of the mitotic chromosome have been proposed, including the radial loop model \cite{Mae03}, the chromatin network model \cite{Poi02} and the hierarchical folding model \cite{Kir04}. These models often highlight the biologically specific role of protein molecules but the intrinsic properties of the DNA as a long highly helical molecule must also be at work. Being a helix itself, DNA, when condensed, forms a range of distinct liquid crystalline phases \cite{Str88}. The DNA of dinoflagellates, which is much longer than a human chromosome, organizes into a cholesteric liquid crystal \cite{Liv78, Gau86}. Human mitotic chromosomes also appear to have liquid crystalline features when viewed under the microscope \cite{de88}. Chiral symmetry appears to be broken: the daughter chromatids formed on cell division seem to be of opposite handedness in microscope images. In this letter, we use inverse statistical mechanics to infer from high resolution chromosome conformation capture data two energy landscapes that yield ensembles of three-dimensional (3D) structures for the mitotic chromosome. One of these energy landscapes is constructed to be agnostic as to the origin of the fluctuating order, while the other is based on a model constructed by adding to an ideal homogenized chromosome landscape that leads to helical order, sequence dependent interactions that give rise to topologically associating domains (TADs) in the interphase chromosome. We use these landscapes to investigate the determinants of the overall chromosome shape, along with the liquid crystalline and chiral ordering by computing the chromosome's phase diagram.

The experimental input for our investigation comes from genome-wide chromosome conformation capture (HiC) studies \cite{Lie09, Dek13}. Via chemical cross-linking, HiC experiments determine the frequency at which any given pair of genomic loci comes close together inside the cell nucleus.  Current experiments resolve the loci to the kilobase level \cite{Rao14}. We use an inverse statistical mechanics algorithm to derive energy landscapes using this pair contact information. Simulating the resulting energy landscapes allows a thorough sampling of 3D conformations of the genome that reproduces the experimental input. Previously we constructed such an effective energy landscape for the interphase chromosome. Because interphase chromatin is not particularly dense, the chromosome at interphase has only locally fluctuating structural order \cite{Zha15}, while the same approach applied to the much denser mitotic chromosome, gives rise to several kinds of broken symmetry: the mitotic chromosome is clearly anisotropic and spontaneously forms cylindrical structures. These structures are twisted and possess a handedness, so chiral symmetry is broken too.

\begin{figure*}[t]
    \centering
    \includegraphics[width=120mm]{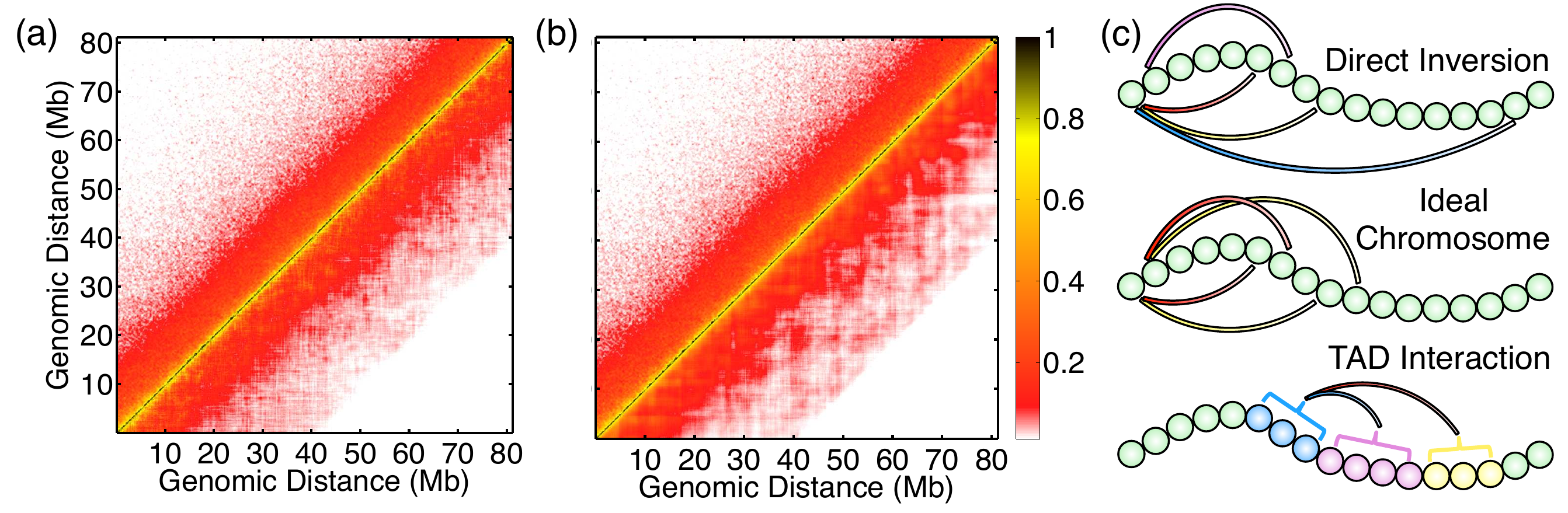}
    \caption{\label{fig:grn}
Comparison between contact probabilities obtained from experiment and simulation. 
(a, b) Contact probability maps for the mitotic chromosome as determined in Hi-C experiments \cite{Nau13} (upper triangles) and as sampled in simulations based on the maximum entropy information theoretic energy landscape (lower triangles) from direct inversion (a) and from the TAD-augmented ideal chromosome model (b). 
(c) Schematic representation for the Hamiltonian for direct inversion (top), for the ideal chromosome model (middle), and for the sequence dependent interactions among topologically associating domains. The chromosome is drawn as beads on a string, and the interactions among genomic loci are drawn as curved lines, with different colors indicating varying strengths. 
    }
\end{figure*}

We model the chromosome at an extremely coarse resolution as beads on a string. Each bead represents 100-kilo base pairs, and is itself a large polymeric object. In the direct inversion, the potential energy function at this coarse bead resolution consistent with maximum entropy is taken to have the form 
\begin{equation}
    U_\mathrm{ME}(r) = U(r) + \sum_{ij} \alpha_{ij}f(r_{ij}).
\end{equation}
$U(r)$ denotes the potential energy function of a connected homopolymer, while $f(r_{ij})$ monitors contact formation. $f(r_{ij})$ represents the probability two polymers whose centers are separated by a distance $r_{ij}$ apart will form a detectable cross link. We take $f(r_{ij})$ to have the form of a switching function inspired by polymer physics for each bead (15). The set of $\{\alpha_{ij}\}$ determine the strength of the contact interactions between specific genomic loci. The $\boldsymbol{\alpha}$ values are found by minimizing an the objective function $\Gamma(\boldsymbol{\alpha})$ defined as  
\begin{equation}
    \Gamma(\boldsymbol{\alpha}) = \ln\frac{Z(\boldsymbol{\alpha})}{Z_o} + \beta \sum_{ij}\alpha_{ij} f_{ij}^{\mathrm{exp}},
\end{equation}
where $\beta$ is the inverse information theoretic temperature. $Z(\boldsymbol{\alpha})$ and $Z_o$ are the partition functions for $U_{\mathrm{ME}} (r)$ and $U(r)$ respectively. Since $Z(\boldsymbol{\alpha})$ is the partition function for the maximum entropy Hamiltonian $U_\mathrm{ME}(r)$ with parameters $\{\alpha_{ij}\}$, $\Gamma(\boldsymbol{\alpha})$ is the loss of information theoretic entropy of an ensemble owing to its being biased to reproduce the input data, having only made the prior assumption that the chromosome is a connected polymer chain with partition function $Z_o$ \cite{Rou13, Pit12}. Using a cumulant expansion, $\Gamma(\boldsymbol{\alpha})$ can be approximated as $\frac{\beta^2}{2}\boldsymbol{\alpha}^T\boldsymbol{B\alpha} - \beta\left[\left<\boldsymbol{f}\right>-\boldsymbol{f}^\mathrm{exp}\right]^T\boldsymbol{\alpha}$, where $\boldsymbol{B}=\left<\boldsymbol{ff}^T\right>-\left<\boldsymbol{f}\right>\left<\boldsymbol{f}^T\right>$ \cite{Eas02}. $\boldsymbol{f}$ and $\boldsymbol{f}^\mathrm{exp}$ are column vectors of contact probabilities for all the pairs of genomic loci included in Eq.\@ [1] determined from the simulation and experimental measurements respectively. The approximate expression for $\Gamma(\boldsymbol{\alpha})$ takes its extreme value at $\boldsymbol{\alpha}=\frac{1}{\beta}\boldsymbol{B}^{-1}[\left<\boldsymbol{f}\right>-\boldsymbol{f}^\mathrm{exp}]$. Since this expression is only an approximate solution for $\boldsymbol{\alpha}$, as outlined in Ref. \cite{Zha15}, we iterate the procedure to determine final values for $\boldsymbol{\alpha}$ for which the contact probabilities determined from simulation agree as closely as possible with the experimental input. A schematic representation of the Hamiltonian from direct inversion is provided in the top panel of Figure 1(c). Because the $\{\alpha_{ij}\}$ are allowed to vary independently, the interactions between different loci adopt distinct values indicated with different colors.

The inferred ``agnostic'' information theoretic energy landscape reproduces the experimental contact probabilities as seen in Figure 1, which compares the experimental and simulated contact probability maps shown in the upper and the lower triangle respectively. As seen in Figure S1 (a), the individual pair probabilities obtained by simulating the landscape are very accurately reproduced with a very high R-squared $\sim 0.93$. 

While using only probabilistic contact pair input alone, the inferred ensemble of structures immediately reproduces the large scale shape and size of chromosome as seen in the microscope. Figure 2(a) shows that the simulated structures naturally exhibit cylindrical shapes, like those seen by light microscopy (1). This observed anisotropy indicates the metaphase chromosome must break rotational symmetry and is not an isotropic fluid or simple collapsed polymer phase. To quantitatively characterize the geometry of the simulated chromosome ensemble, we determined the statistics of extension lengths of each structure along its three principal axes, as shown in the inset of Figure 2(a). The distribution for the longest axes peaks at a value twice that for the shorter two, which are comparable, indicating the typical cylindrical shape of the mitotic chromosome. In contrast, when a landscape for the interphase chromosome is constructed using HiC data synchronized at the G1 phase, the resulting structures are more nearly spherical with principal axis ratios typical of an isotropic weakly confined polymer (see Figure S2 (f)). 

\begin{figure}[t]
    \centering
    \includegraphics[width=80mm]{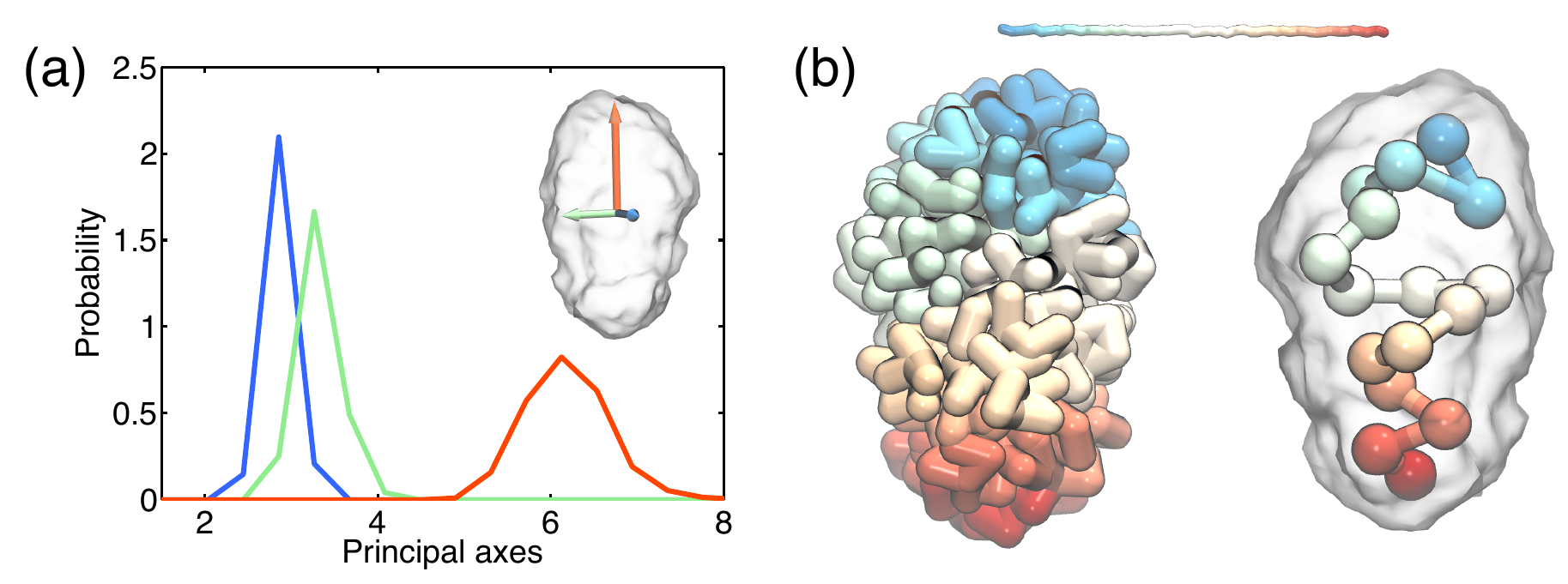}
    \caption{\label{fig:grn}
Structural characterization of the simulated mitotic chromosome conformations.  
(a) Probability distributions for the extension lengths along the three principal axes, an illustration of which is provided in the inset. A sampled chromosome structure is shown in transparent surface in the inset along with the principal axes indicated as arrows.  
(b) A representative structure from the simulated ensemble shown at full resolution (left) and at coarsened resolution (right) illustrates the presence of twist along the sequence. For the coarsened picture, the full resolution chromosome is shown in transparent surface. A color bar that indicates the color variation along the sequence is provided at the top. 
    }
\end{figure}

Figure 2(b) shows a representative simulated structure of the mitotic chromosome. The bead color traverses from blue to cyan to yellow and to red along the sequence. The structure shown in the left at full resolution suggests the presence of a helical twist along the sequence, which is easier to visualize in the coarsened representation shown on the right for segments 5Mb in length, corresponding to a 50 times coarser description. The global helix-like structure clearly signals the presence of cholesteric liquid crystalline ordering in the mitotic chromosome. Helical conformations have indeed been observed experimentally for the mitotic chromosome (11), and sister chromatids produced on cell division are reported to break chiral symmetry, each having an opposite helical handedness. 

To characterize quantitatively the magnitude of the helical twist, we introduce a local collective chiral variable $\psi(i)$ along the chromosome as illustrated in the inset of Figure 3(a), modeled on chirality variables introduced for peptides and nucleic acid chains \cite{Mor09, Mor13}. For a given genomic locus $i$, $\psi$ is defined using four genomic positions $\{i,i+\frac{1}{2} T,i+\frac{3}{4} T,i+\frac{5}{4} T\}$, denoted $A,B,C$ and $D$ respectively, as $\psi(i) = \frac{\overrightarrow{EF}\cdot (\overrightarrow{CD} \times \overrightarrow{AB} )}{|\overrightarrow{EF}|\cdot|\overrightarrow{CD}|\cdot|\overrightarrow{AB}|}$,
where $E$ and $F$ are the midpoints of the vectors $\overrightarrow{AB}$ and $\overrightarrow{CD}$ respectively. We chose the sequence separation $T=25 $Mb to correspond to the period of the large scale helical twist observed in chromosome structures (See Figure 2(b)). $\psi(i)$ distinguishes right-handed from left-handed twists with positive and negative values respectively. 

Figure 3(b) plots the $\psi(i)$ for the ensemble of simulated chromosome structures with red corresponding to a right-handed twist and blue left-handed. Each column in this figure corresponds to a single sampled chromosome structure. A full chirally ordered structures would appear as a uniformly red or blue column. Most columns, however, break into blocks of red and blue segments, indicating that global ordering is often kinetically or thermodynamically prevented from going to completion. As shown in Figure S1 (f), the genomic distance correlation length for the chirality variable $\psi(i)$ is shorter than the entire chromosome. So the probability distribution of $\psi(i)$ averaged over the entire chromosome has only a single peak near zero (Figure 3(a) black), due to the cancellation from a mixture of right-handed (positive) and left-handed (negative) twists in still rather long segments. 

\begin{figure}[t]
    \centering
    \includegraphics[width=80mm]{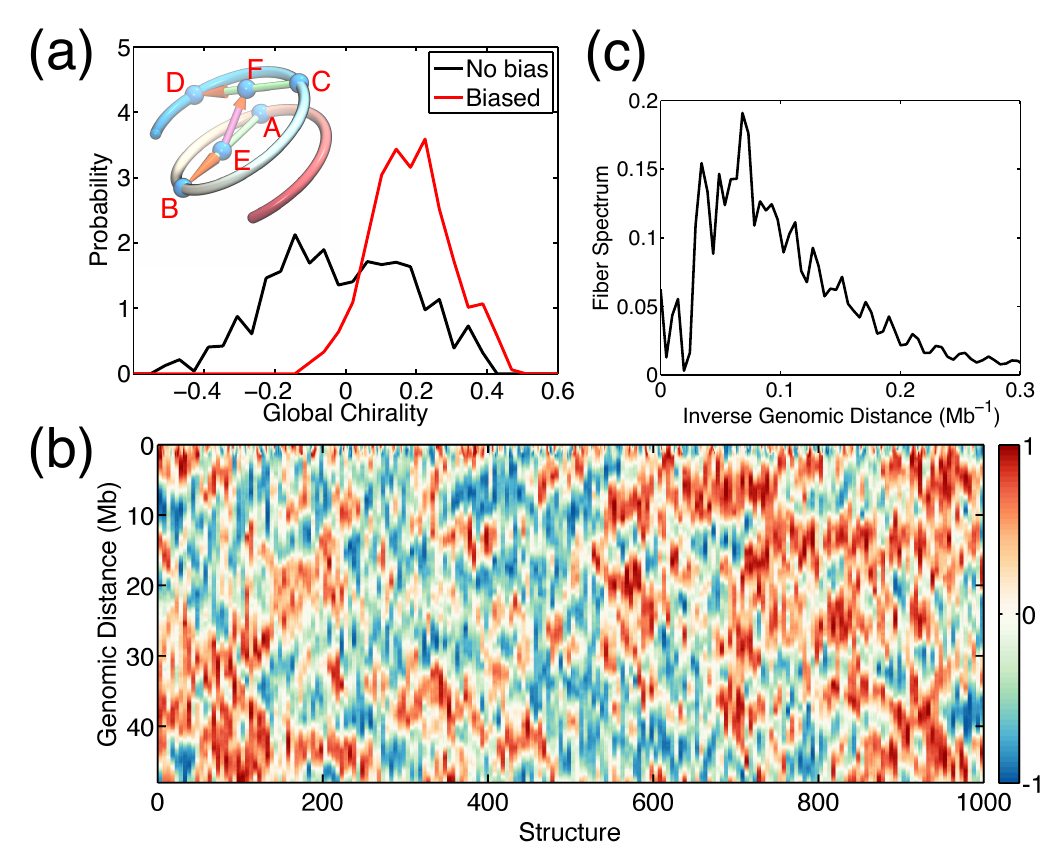}
    \caption{\label{fig:grn}
Quantitative characterization of twist motifs in chromosome structures.   
(a) The probability distributions of the global chirality for the ensemble of structures from chromosome models without torsional bias (black) compared with that with torsional bias (red). An illustration defining of the collective variable $\psi$ that measures the handedness of a given chiral structure is shown in the inset. A segment of the chromosome is drawn as a cylindrical shape, and the three arrows correspond to the vectors used to define $\psi$. 
(b) Density plots of $\psi$ along the genomic sequence for different chromosome structures from the simulated ensemble. 
(c) Fiber spectrum of the metaphase chromosome structures simulated with the presence of weak torsional biases.
    }
\end{figure}

Clearly the chromosome has a high susceptibility to chiral order. Since the chiral domain boundaries can at best diffuse, we believe kinetics prevents observing complete global symmetry breaking. To test for global ordering, we then explicitly but weakly break the chiral symmetry by applying a weak torsional bias locally to the chain. As shown in Figure S1 (e), the distribution of torsional dihedral angles without the weak bias exhibits two peaks at both positive and negative values, which arise from the right-handed and left-handed segments in the structure respectively. In contrast, as shown in Figure S3, when weak biases are introduced, reoptimization leads to a unimodal distribution. The resulting structures now exhibit a global twist, with the chirality distribution peaking around 0.2 (Figure 3(a) red line). Remarkably, the contact maps obtained from simulations incorporating the weak local dihedral bias are equivalent in quality to those found without introducing any local chirality bias (See Figure S3).
	
The HiC contact map data by themselves are not sufficient to distinguish structures with global twist from an ensemble of structures in which the chromosome is broken into long segments with differing chirality. Since the HiC data contain only pair-wise information, mirror image conformations are energetically degenerate. The presence of a global chiral twist of the mitotic chromosome structure is consistent with what is postulated in hierarchical models, in which the chromosome folds successively by forming fibers of fibers \cite{Kir04}. Hierarchical models incorporate multiple layers of twisting and fibril structures at several different lengthscales. The chiral twist collective variable probes structure at the 25 Mb scale. The fiber spectrum which we introduced in our previous study of the interphase chromosome gives evidence for structures at finer scales \cite{Zha15}. The fiber spectrum is the sequence Fourier transform of an orientation order parameter defined by the scalar product of two displacement vectors along the genomic sequence. The fiber spectrum exhibits peaks at frequencies (inverse sequence separation) that correspond to the period of the twist. The fiber spectrum in Figure 3(c) reveals additional helical structures at shorter lengthscales for the mitotic chromosome.  

A hierarchically organized chromosome with liquid crystalline features and orientational ordering can arise from a generic ideal chromosome model \cite{Zha15}. An ideal chromosome potential represents an effective homopolymer with a pair-wise contact potential, whose strength depends only on the sequence separation of the two genomic loci, not on their specific locations. This homogenized model encodes only the sequence translation invariant features of the DNA along with its protein packaging. An illustration of the ideal chromosome model is provided in the middle panel of Figure 1(c). The interphase chromosome displays clear deviations from the best equivalent homopolymer landscape, exhibiting clumps or topologically associating domains that are of size $\sim 1$ Mb along the genomic sequence \cite{Dix12}, as highlighted in the interphase HiC map shown in Figure S5. Topologically associating domains have different histone modifications \cite{Rao14, Sex12}, and may play crucial roles in gene regulation \cite{Zha14, Le14, Che15}. Establishing strict boundaries between topologically associating domains necessitates there being deviations from the homogenized ideal chromosome model. These energetic inhomogeneities can pin the motions of defects in helical or chiral ordering \cite{Lut95}. Sequence specific interactions among topologically associating domains likely arise from the varying chemical compositions of domains. Here we study whether these same fluctuations that form topologically associating domains in the interphase persist in modifying the structure and dynamics of metaphase chromosomes. 

To systematically investigate the interplay of the homogeneous ideal chromosome interaction and the sequence specific interactions in organizing and condensing the mitotic chromosome, we introduce a model based on those features of the interphase state. The Hamiltonian for this TAD-augmented ideal chromosome model is 
\begin{eqnarray}
    \nonumber
    U_\mathrm{TI}(r) = U(r) &+& \sum_{ij}\alpha_\mathrm{ideal}(|j-i|)f(r_{ij}) \\
                            &+& \lambda\sum_{A,B}\alpha_{AB} \sum_{i\in A}\sum_{j\in B}f(r_{ij})
\end{eqnarray}
The first term is the same homopolymer potential as in Eq.\@ [1]. The second term corresponds to the homogeneous ideal chromosome contact potential, whose strength depends only on the sequence separation $|j-i|$ and is therefore sequence translation invariant. The final term explicitly breaks sequence translation invariance and describes sequence specific interactions at the level of the topologically associating domains found in the interphase, as illustrated in the bottom panel of Figure 1(c). All the genomic loci in a given topologically associating domain A are treated equally and, when interacting with loci from another domain B, they share the same contact potential $\alpha_{AB}$. The prefactor $\lambda$ of the third term is introduced for studying the consequences of the pinning terms, and is set to $1$ during the iterative fitting of the experimental contact map. 

Again we use the maximum entropy scheme as in Eq.\@ [3] to find the parameters for both the interphase and metaphase chromosomes. For the metaphase chromosome, we also further incorporated very weak dihedral biases as in the direct inversion to encourage global twisting. Figure 1(b) and Figure S6 together demonstrate that the TAD-augmented ideal chromosome model of the mitotic chromosome again reproduces the contact map and the power law scaling of the contact probability. Just as for the direct inversion, chromosome conformations produced with $U_\mathrm{TI} (r)$ adopt cylindrical shapes with global twisting. As shown in Figure S6 (f), the energies for the TAD-augmented ideal chromosome model are significantly correlated to those from the Hamiltonian obtained using direct inversion when sampled over the structural ensemble. 

\begin{figure}[t]
    \centering
    \includegraphics[width=80mm]{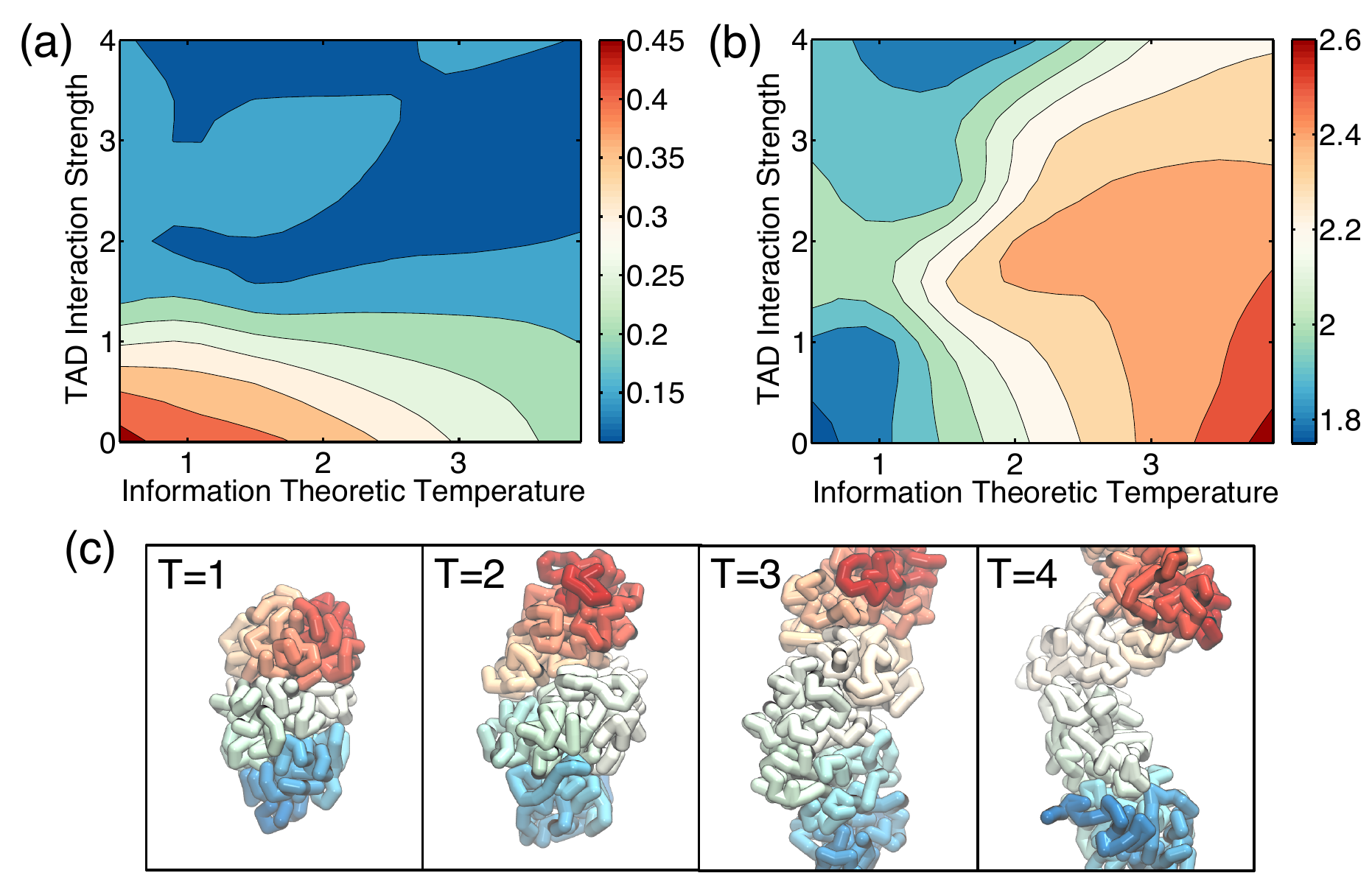}
    \caption{\label{fig:grn}
Phase diagrams for the TAD-augmented ideal chromosome model of the mitotic chromosome.   
(a, b) Global chirality (a) and ratio of the largest and smallest extension length along the principal axes (b) are plotted as a function of temperature $T$ and inhomogeneity strength $\lambda$. 
(c) Representative chromosome structures at varying temperature $T$ that illustrate the loss of chirality as the helical twist unwinds. 
    }
\end{figure}

Comparing the interphase and metaphase potentials suggests a possible mechanism for chromosome condensation. As shown in Figure S8 (a), the main difference between the interphase and metaphase ideal chromosome contact potentials $\alpha_\mathrm{ideal} (|j-i|)$ is that the metaphase interactions are consistently stronger at both short and long range sequence separation. Even when the domain inducing interactions are neglected, the resulting ideal chromosome landscape including only the first two terms in Eq.\@ [3], gives rise to condensed and twisted structures as shown by the chirality variable and the length of the principal axes in Figure S8. In the main the condensation of the mitotic chromosome can be understood through the emergence of a strong ideal chromosome potential, that naturally leads to hierarchical fiber ordering.

To further understand the role of the ``random'' fields that arise from interactions among topologically associating domains that form in the interphase chromosome, we compute the phase diagram for $U_\mathrm{TI}(r)$ as a function of both temperature $T$ and the inhomogeneity prefactor $\lambda$. Two dimensional plots of the global chiral ordering and the ratio of the lengths for the longest and shortest principal axes are presented in Figures 5(a) and 5(b) respectively. While the homogeneous ideal chromosome contact potential promotes both hierarchical fiber and chiral ordering, the interactions among topologically associating domains destroy chiral order. Chiral ordering is also lost with increasing temperature $T$ through the unwrapping of the helical twist. This is evidenced by the increase of chromosome length shown in Figure 5(b). Examples of chromosome conformations at various temperatures for $\lambda=1$ are provided in Figure 5(c). On the other hand, the chromosome still remains condensed at large inhomogeneity strength $\lambda$, and the chiral ordering is lost through the formation of defects in the chiral structure.

\begin{figure}[t]
    \centering
    \includegraphics[width=80mm]{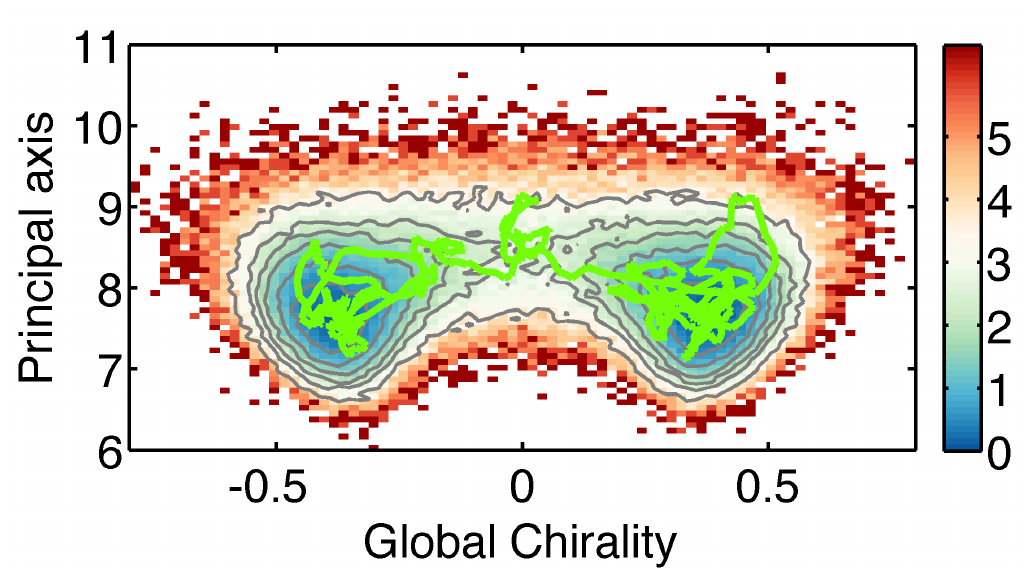}
    \caption{\label{fig:grn}
Coupling between chromosome shape changes and chiral ordering.  The free energy as a function of the global chirality and the length of the longest principal axis is shown as a surface plot. Contours of constant free energy are drawn as grey lines. An example trajectory that connects the two chiral conformations is plotted in green. 
}
\end{figure}

Chromosome macroscale organization arises from a balance between the ideal chromosome contact potential that promotes helical, chiral, and cylindrical ordering and the interactions among topologically associating domains that encourage defects in such order. The coupling between shape changes and the chiral and liquid crystalline ordering is strong. We can see this clearly through equilibrium simulations using the homogenized ideal chromosome model without torsional bias at a high information temperature $T=1.5$. The combination of high temperature and the absence of defect pinning interactions allows full equilibration. The chirality transition in this ideal model is accompanied by the structural extension of the chromosome. This can be seen in Figure 6, which displays the free energy surface as a simultaneous function of the longest principal axis length of the chromosome and its global chirality. An example reactive trajectory shown in green plotted on top of this free energy surface shows that extending of the chromosome can switch the chirality, suggesting the chiral order in natural chromosomes may be setup by the cellular machinery that locates the chromosome within the cell.

This work was supported by the Center for Theoretical Biological Physics sponsored by the NSF (Grants PHY-1308264 and PHY-1427654). P.G.W. acknowledges financial support by the D. R. Bullard-Welch Chair (Grant C-0016) at Rice University.

%

\end{document}